\documentclass[aps,prl,twocolumn,groupedaddress,showpacs]{revtex4}
\usepackage{amssymb}
\usepackage{graphicx,color}
\definecolor{r}{rgb}{1,0,0}   
\definecolor{g}{rgb}{0,1,0}   
\definecolor{b}{rgb}{0,0,1}

\begin{document}


\title{Microfluidic rheology of soft colloids above and below jamming}


\author{K. N. Nordstrom$^1$, E. Verneuil$^{1,2}$, P. E. Arratia$^{1,3}$, A. Basu$^1$, Z. Zhang$^{1,2}$, A. G. Yodh$^1$, J. P. Gollub$^{1,4}$ and D. J. Durian$^1$}
\affiliation{$^1$Department of Physics and Astronomy, University of Pennsylvania, Philadelphia, PA 19104, USA}
\affiliation{$^2$Complex Assemblies of Soft Matter, CNRS-Rhodia-UPenn UMI 3254, Bristol, PA 19007, USA}
\affiliation{$^3$Department of Mechanical Engineering, University of Pennsylvania, Philadelphia, PA 19104, USA}
\affiliation{$^4$Department of Physics and Astronomy, Haverford College, Haverford, PA 19041, USA}


\date{\today}

\begin{abstract}
The rheology near jamming of a suspension of soft colloidal spheres is studied using a custom microfluidic rheometer that provides stress versus strain rate over many decades. We find non-Newtonian behavior below the jamming concentration and yield stress behavior above it. The data may be collapsed onto two branches with critical scaling exponents that agree with expectations based on Hertzian contacts and viscous drag. These results support the conclusion that jamming is similar to a critical phase transition, but with interaction-dependent exponents.
\end{abstract}

%


\maketitle




The concept of jamming \cite{LiuNagelBOOK} aims for a unified understanding of dense soft matter such as granular media, colloids, foams, emulsions, and glassy liquids.  Such systems are considered to be jammed if the relaxation time is longer than the observation window, so that the particles appear stuck in a fixed configuration and the sample appears to have a yield stress.  At zero temperature and zero applied stress, this transition occurs at a critical volume fraction of particles, $\phi_c$, and is known as ``point J'' on the jamming phase diagram \cite{OHernPRE2003}.  The discovery of growing dynamical length and time scales on approach to point J suggests that it is a critical point \cite{DJDPRL1995,OHernPRE2003, EllenbroekPRL2006, AbateNatPhys2007, ZhangNature2009}.

To extend these ideas to driven systems, the shear stress $\sigma$ has been studied theoretically as a function of both imposed strain rate $\dot\gamma$ and of distance $|\phi-\phi_c|$ to the critical packing fraction \cite{OlssonPRL2007, HatanoJPSJ2008, OtsukiPRE09, Tighe2010}.  The rheology was found to collapse onto separate branches above and below $\phi_c$ when plotted as $\sigma/|\phi-\phi_c|^\Delta$ versus $\dot\gamma/|\phi-\phi_c|^\Gamma$.  For Hertzian particles in a viscous liquid, conflicting exponents have been reported as \{$\Delta=1.8\pm0.1$, $\Gamma=2.4\pm0.1$\} \cite{HatanoJPSJ2008},  \{$\Delta=1.5$, $\Gamma=2.75$\} \cite{OtsukiPRE09}, and \{$\Delta=2$, $\Gamma=4$\} \cite{Tighe2010}. However there has been no experimental test of these predictions. In part this is because most experiments are restricted to hard spheres, which cannot be packed above $\phi_c\approx0.64$.  Also, conventional rheometers can give misleading results for yield-stress materials due to wall-slip and shear banding.  Here, we circumvent both issues by using soft gel particles and a custom microfluidic rheology technique.  We demonstrate critical behavior by the collapse of scaled stress versus strain rate, with exponents that can be understood in terms of particle-particle interaction forces in best agreement with Ref.~\cite{Tighe2010}.

\begin{figure}
\includegraphics[width=3.00in]{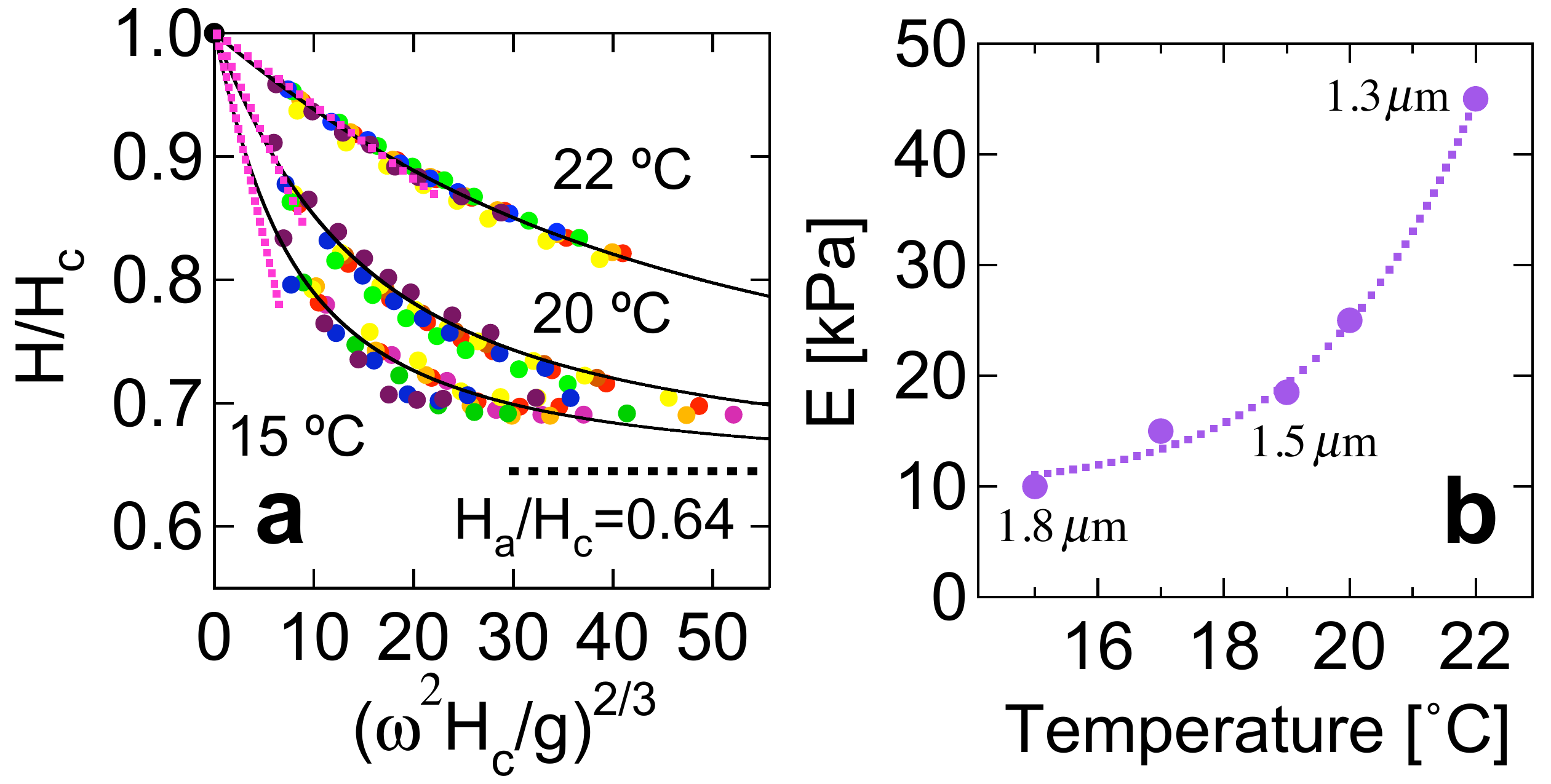}
\caption{(Color online)
Characterization of NIPA particles. (a) Normalized height of a column of particles vs dimensionless combination of angular rotation speed $\omega$, initial height $H_c$, and $g=9.8~{\rm m/s}^2$; colors distinguish different $H_c$.    (b) Elastic modulus of the particles vs temperature, deduced from the initial slopes shown as dashed lines in (a), plus empirical fit corresponding to $E \sim 1/V^3$ where $V$ is particle volume.  Particle diameters are indicated along the curve.} \label{NIPA}
\end{figure}

The experimental system is a dense aqueous suspension of colloidal N-isopropylacrylamide (NIPA) gel particles \cite{SaundersACIS99, PeltonACIS00}, which are swollen with water and hence are nearly index and density matched. By changing the temperature, the amount of swelling and hence the volume fraction can be readily controlled.  As characterized by dynamic light scattering for dilute samples, the diameter is of order 1~$\mu$m and the volume varies with temperature as  $V(T)=(2.93~\mu{\rm m}^3)[1-T/(39.6\ ^\circ{\rm C})]$ for our temperature range, $19\ ^\circ $C$\leq T\leq25\ ^\circ $C$ $. The polydispersity is about ten percent. Here we study a single sample with particle number density of $4.55\times10^{17}/$m$^3$, such that the empirical fit translates to $\phi(T)=1.34-T/(29.4\ ^\circ $C$)$ for our temperature range.  

The mechanical properties of the particles are characterized by centrifugal compression; see Ref.~\cite{CentComp} for full details.  At rest, the particles sediment to a total packing height $H_c$ corresponding to random close packing of spheres at $\phi_c\approx0.64$.  When spun at angular speed $\omega$, the packing compresses to a smaller height $H$, and the volume fraction varies with depth such elastic and centrifugal forces balance everywhere.  Data for $H/H_c$ are plotted vs $(\omega^2 H_c / g)^{2/3}$ in Fig.~\ref{NIPA}a, where $g=9.8$~m/s$^2$, for three different temperatures and for samples with a range of $H_c$ values.  The choice of x-axis is both so that the data collapse for different $H_c$, and so that the initial decay is linear for Hertzian spheres and can be used to deduce the elastic modulus $E$ of the sphere material \cite{CentComp}.   This holds well; final results for $E$ are plotted vs temperature in Fig.~\ref{NIPA}b, and will used to scale rheology data. The solid black curves in Fig.~\ref{NIPA}a represent fits based on a constitutive elastic model where the compressive stress is Hertzian at small strains and diverges at a finite strain \cite{CentComp}.  The striking feature is that these fits all asymptote to $H_a/H_c\approx0.64$ at high centrifugal acceleration.  Since the average volume fraction is inversely proportional to height, then $H_a/H_c=\phi_c/\phi_a$.  The asymptotic normalized compression in Fig.~\ref{NIPA}a therefore gives an asymptotic packing fraction of $\phi_a=1$.  This suggests that the NIPA particles deform under compression without deswelling, so that packing fractions {\it above} $\phi_c$ also may be reliably computed from the dilute suspension particle size data.

\begin{figure}
\includegraphics[width=3.0in]{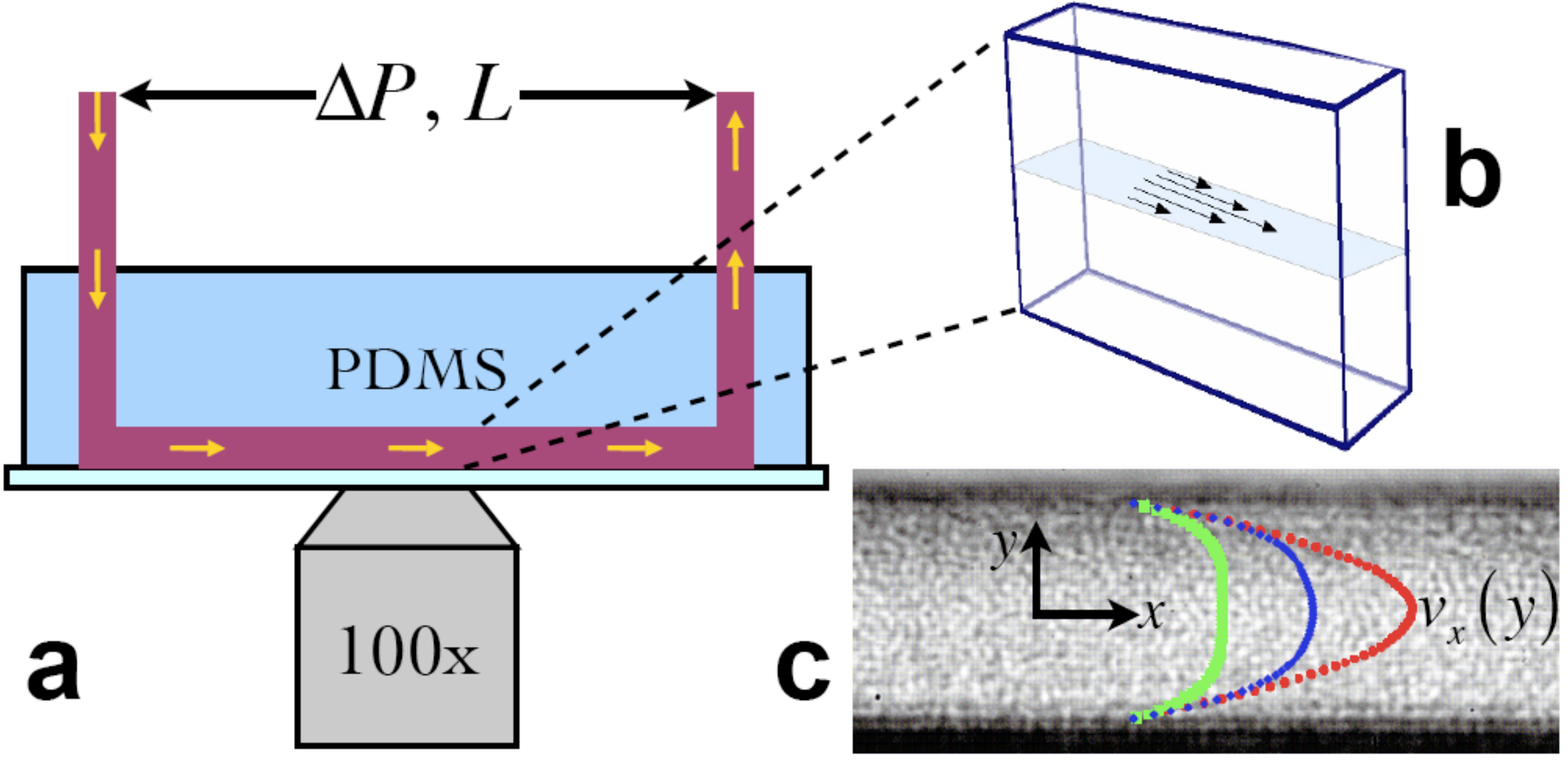}
\caption{(Color online) The experimental setup. (a) The fluid is driven through the microfluidic channel by setting the inlet and outlet pressures. (b) Video data of the suspension flow is taken at mid-channel height. (c) An image taken from a video showing the particles. Example velocity profiles are superimposed for a Newtonian fluid (red) and for NIPA samples at $\phi=0.56$ (blue) and $\phi=0.64$ (green).} 
\label{SYSTEM} 
\end{figure}

The steady shear flow rheology is measured with the microfluidic device sketched in Fig.~\ref{SYSTEM}a.  It consists of a rectangular PDMS channel, 25~$\mu$m wide $\times$ 100~$\mu$m deep $\times$ $L=2$~cm long, fabricated with standard soft lithography \cite{XiaARMS1998} and bonded to a glass microscope slide.  The fluid is forced through the channel using pressurized air and inlet/outlet tubing of sufficient diameter that the imposed pressure drop $\Delta P$ occurs only along the length $L$ of the channel. Force balance therefore allows the shear stress at distance $y$ from the center of the channel to be computed as $\sigma(y) = (\Delta P/L)y$.  The corresponding strain rate at $y$ is found by numerical differentiation of the velocity profile, $\dot\gamma(y)={\rm d}v_x(y)/{\rm d}y$ \cite{SOOM}. For this, we collect video data with a Phantom CMOS camera (1-10,000~fps) connected to a Zeiss Axiovert 200 microscope with $100\times$ objective focussed at mid-height.  Since the channel is tall, the observed flow is equivalent to that between parallel plates \cite{SOOM}.  An objective-cooling collar (Bioptechs) and cooling plate above the sample are controlled to about 0.1~C in order to vary the volume fraction.  An example video frame in Fig.~\ref{SYSTEM}c displays bead-scale intensity variations, so that Particle Image Velocimetry may be implemented with custom LabVIEW code. Example velocity profiles are superposed on the still image of Fig.~\ref{SYSTEM}c.  Altogether, for a single pressure drop, the $\sigma(y)$ and $\dot\gamma(y)$ data may thus be combined to give stress vs strain rate shear rheology.  The dynamic range is typically two decades in $\dot\gamma$, and may be extended by varying the imposed pressure drop.

This rheology concept has been realized previously \cite{DegreAPL2006, GoyonNature2008}, and is related to experiments \cite{DanPRL94, KatgertPRL08, KatgertEPL2010} where the shape of a velocity profile is used to characterize shear rheology.  Microfluidics is an ideal platform, since the channels are long compared to width so that entrance/exit effects are easily avoided.  And owing to the small scale, high strain rates may be achieved at low Reynolds numbers, so that inertial flow instabilities are avoided. Furthermore, the local strain rate is directly measured, and hence no problems arise from wall slip or shear banding as typically hamper use of conventional rheometers for materials with a yield stress.

\begin{figure}
\includegraphics[width=3.0in]{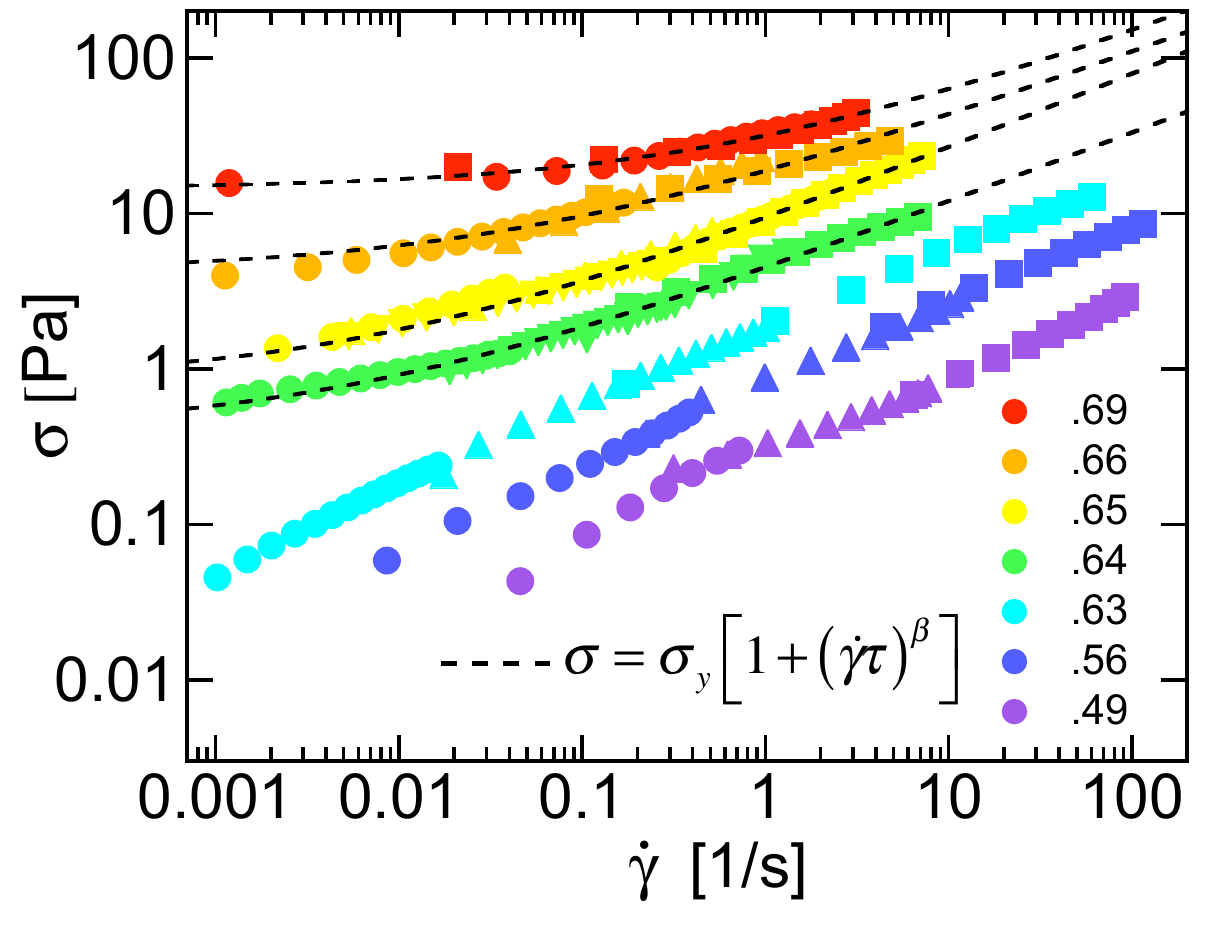}
\caption{(Color online) Shear stress versus strain rate for several different volume fractions, as labeled.  Symbol types distinguish runs at different driving pressures. The dashed curves represent fits to the Herschel-Bulkley equation.}
\label{data}
\end{figure}

\begin{figure}
\includegraphics[width=3.0in]{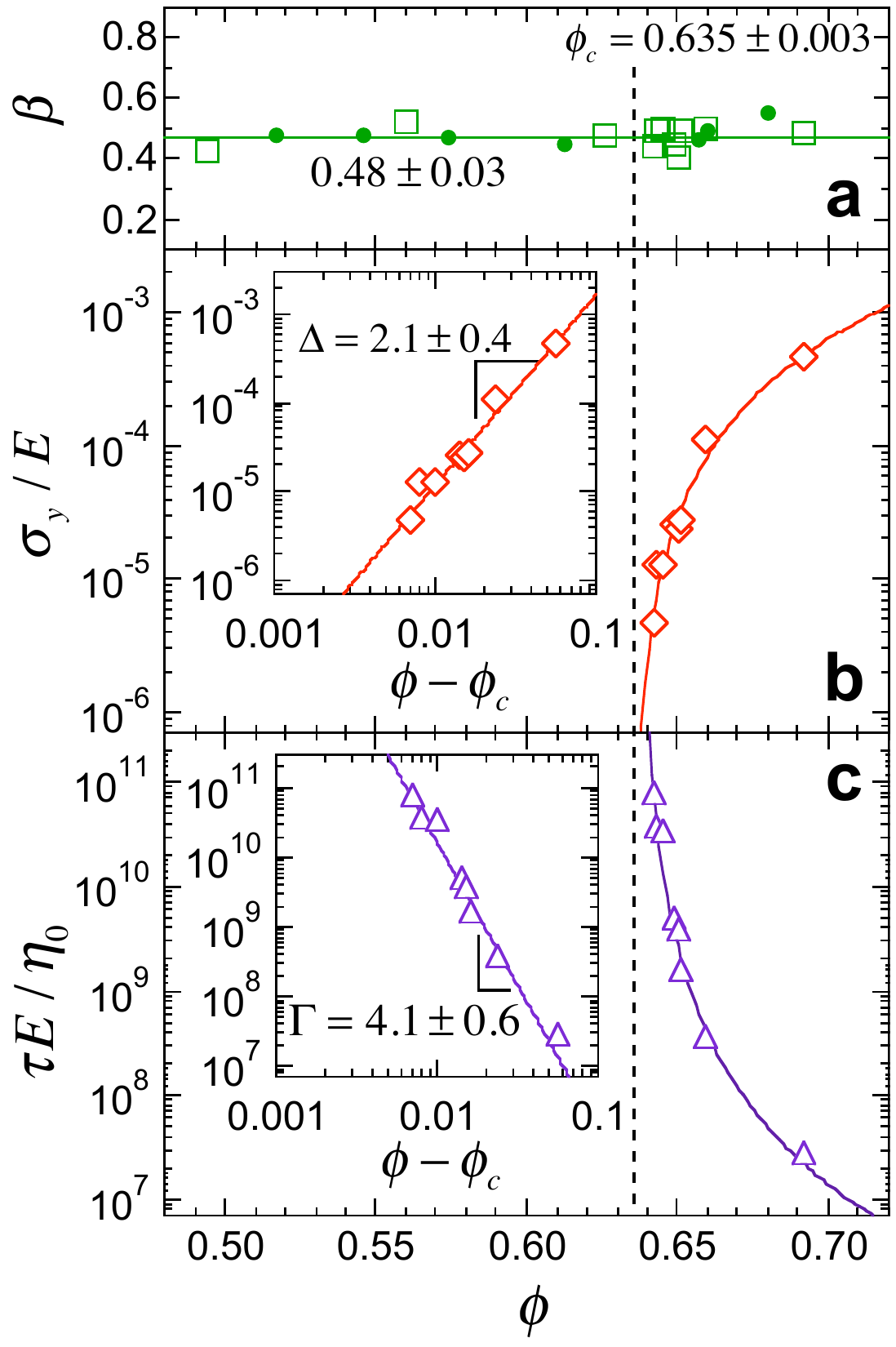}
\caption{(Color online) Fitting parameters vs volume fraction $\phi$: (a) the exponent $\beta$, (b) the dimensionless yield stress $\sigma_y/E$, and (c) the dimensionless timescale $\tau E/\eta_0$; $E$ is the particulate material elastic modulus and $\eta_0=0.01$~g/(cm-s) is the fluid viscosity.  The large symbols are for the main system of particles, as in Figs.~1 and 3; the small symbols in part (a) are for particles about 8 times less massive.}
\label{stackplot}
\end{figure}

Results for stress vs strain rate are collected in Fig.~\ref{data}.  Good agreement is found for multiple pressure drops at the same volume fraction.  This demonstrates the reproducibility and level of uncertainty in our data; it also implies the absence of non-local effects, by contrast with Refs.~\cite{GoyonNature2008, KatgertEPL2010}. Note that the data show a clear distinction in functional form above and below $\phi_c$.  For low $\phi$, the stress tends towards zero at low strain rates. For higher $\phi$, the stress extrapolates toward a non-zero yield stress $\sigma_y$. We find qualitatively similar results using a conventional rheometer.  To analyze the flow curves, we first fit the stress data to the phenomenological Herschel-Bulkley form:
\begin{equation}
  \sigma = \sigma_y[ 1 + (\dot\gamma \tau)^\beta] = \sigma_y + K{\dot\gamma}^\beta,
\label{HB}
\end{equation}
where $\beta$ is the shear-thinning exponent, $\tau$ is a time constant, and $K$ is called the consistency.  The quality of the fits is satisfactory, as shown by the dashed curves in Fig.~\ref{data} for $\phi>\phi_c$.  The results for $\beta$ displayed in Fig.~\ref{stackplot}a exhibit no apparent dependence on $\phi$, and have average and standard deviation $0.48\pm0.03$.    For other microgel systems, Ref.~\cite{GraessleyRA88} discusses yield stress behavior and Ref.~\cite{CloitrePRL03} finds $\beta=0.45$, while Refs.~\cite{WolffeJCIS89, JonesJCIS92, RichteringJCP99, TrappePT09} fit to forms that cross between different limiting viscosities at low and high strain rates.   The value $\beta=1/2$ is predicted near jamming for viscously-interacting athermal particles \cite{Tighe2010}.  For simplicity, and so that $K$ has constant units, we henceforth fix $\beta=1/2$ and repeat the fits.

\begin{figure}
\includegraphics[width=3.0in]{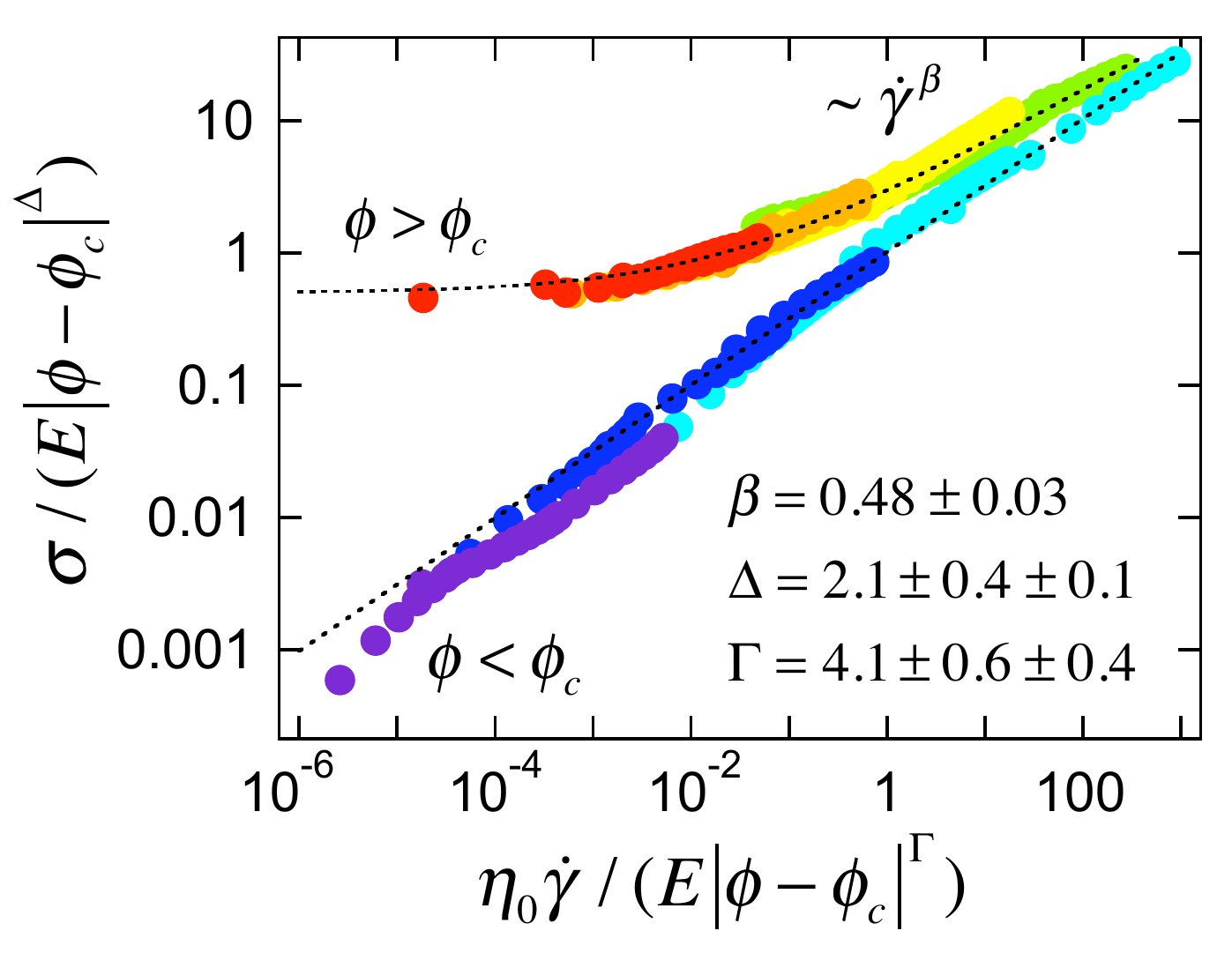}
\caption{(Color online) Collapse of stress vs strain rate using the critical exponents $\Delta$ and $\Gamma$. The dashed lines are fits to
the Herschel-Bulkley form with $\beta=1/2$.  The first exponent uncertainties are statistical; the second are systematic and reflect the allowed range of $\beta$.} 
\label{collapse}
\end{figure}

The fitting parameters $\sigma_y$ and $\tau$ are collected in Fig.~\ref{stackplot}b-c as a function of $\phi$. Both the yield stress and the time constant have been rendered dimensionless by appropriate factors of the elasticity $E$ of the particulate material and the viscosity $\eta_0$ of the suspending fluid. This also serves to eliminate the spurious $\phi$-dependence originating from the variation of $E$ with particle swelling. While $K$ is always well-defined, $\sigma_y$ and $\tau$ exist only above jamming and respectively appear to vanish and diverge on approach to $\phi_c$.  As shown in the main plots, the results may be fitted to power-law forms $\sigma_{y}/E \sim (\phi-\phi_{c})^{\Delta}$ and $\tau E/\eta_0 \sim 1/ (\phi-\phi_{c})^{\Gamma}$, giving \{$\phi_c=0.633\pm0.002$, $\Delta=2.2\pm0.4$\} and \{$\phi_c=0.637\pm0.002$, $\Gamma=3.8\pm0.6$\}.   The two values for $\phi_c$ are in agreement and average to $0.635\pm0.003$, consistent with random-close packing of spheres.  Fixing $\phi_c$ to this value, we plot $\sigma_y/E$ and $\tau E/\eta_0$ vs $\phi-\phi_c$ on log-log axes as insets in Figs.~\ref{stackplot}b-c.  These demonstrate power-law behavior, and give the final refined scaling exponents as $\Delta=2.1\pm0.2$ and $\Gamma=4.1\pm0.3$.  However we will conservatively take the final statistical uncertainties to be twice as large, as given by fits where $\phi_c$ floats.  The systematic errors based on the allowed range of $\beta$ are 0.1 and 0.4 for $\Delta$ and $\Gamma$, respectively.   Note that $\Delta=\beta\Gamma$ holds within uncertainty, which is required so that $K$ remains finite and nonzero at $\phi_c$ and so that at high strain rates the stress scales as $(\eta_0\dot\gamma)^\beta E^{(1-\beta)}$ independent of $\phi$.  Also, the very same exponents are found within experimental uncertainty for NIPA particles about 8 times less massive \cite{SOOM}.

Our experimental value of $\Delta$ agrees with that simulated in Ref.~\cite{HatanoJPSJ2008}, and our full suite of \{$\beta$, $\Delta$, $\Gamma$\} values are in remarkably good agreement with those predicted in Ref.~\cite{Tighe2010}.  The observed value of the yield-stress exponent may be understood physically in terms of the scaling of the shear modulus $G$ and the yield strain $\gamma_y$.  For repulsive particles with interaction energy proportional to overlap raised to the power $\alpha$, numerical simulations find $G \sim (\phi-\phi_c)^{\alpha-3/2}$; this differs from the naive expectation $\alpha-2$ due to $\phi$-dependent non-affine motion \cite{OHernPRE2003, EllenbroekPRL2006}.  If the yield strain scales as $\gamma_y \sim (\phi-\phi_c)$, and if the yield stress scales as $\sigma_y \approx G\gamma_y \sim (\phi-\phi_c)^{\alpha-1/2}$, then $\Delta = \alpha-1/2$ \cite{Tighe2010}.  For Hertzian elastic particles, $\alpha=5/2$, this predicts $\Delta=2$ and $\Gamma=\Delta/\beta=4$ as seen here.

The ``distance'' $\phi-\phi_c$ to jamming thus controls the yield stress $\sigma_y$ and the time constant $\tau$ appearing in the Herschel-Bulkley form of stress vs strain rate, Eq.~(\ref{HB}), according to respective scaling exponents $\Delta$ and $\Gamma$.  Therefore, for volume fractions above $\phi_c$, the shear rheology data should all collapse onto a single master curve when plotted dimensionlessly as $\sigma/(E|\phi-\phi_c|^\Delta)$ vs $\eta_0\dot\gamma/(E|\phi-\phi_c|^\Gamma)$. This construction and the required collapse for $\phi>\phi_c$ are demonstrated in Fig.~\ref{collapse}.  A noteworthy feature of this plot is that collapse {\it also} occurs for all data {\it below} jamming, for $\phi<\phi_c$, onto a distinct branch.  Note that the two branches merge close to where the dimensionless scaled stress and strain rate are both near 1, which is reassuring.  The collapse along a second branch need not have happened, and serves to emphasize that behavior is controlled by distance to point J -- just as second order phase transitions are controlled by distance to criticality.  The quality of the collapse is only slightly better using $\beta=0.48$ rather than $\beta=1/2$ \cite{SOOM}.  As first discussed in Ref.~\cite{OlssonPRL2007}, plots such as Fig.~\ref{collapse} represent the measurement of universal scaling functions; here, the distinct branches are approximately Herchel-Bulkley above $\phi_c$  and approximately power-law below $\phi_c$.  Recently it was argued that collapse may be unique to Hertzian particles~\cite{Tighe2010}.    
 
Before closing we note that an alternative collapse procedure was also explored, using polynomials rather than the Hershel-Bulkley form for stress vs strain rate \cite{SOOM}.  The best collapse to two distinct branches is found for $\Delta = 2.2\pm0.4$, $\Gamma = 5.2\pm0.8$, and $\phi_c = 0.64\pm0.01$.  This is consistent with the previous analysis, marginally so for $\Gamma$, and indicates the extent to which our analysis of scaling behavior is model-independent.

In summary, we have used a custom microfluidic rheometer to obtain reliable stress vs strain rate data for thermoresponsive NIPA particle suspensions above and below jamming, free from shear-banding and wall-slip artifacts.  Furthermore we have characterized the mechanical properties of the particles themselves using centrifugal compression, both to account for the change in elasticity with temperature and to demonstrate that the particles do not deswell when packed above $\phi_c$ and hence have a well-known volume fraction.  Above jamming, we find that the yield stress scales approximately as $(\phi-\phi_c)^2$.  For all volume fractions examined, we find that the stress increases with strain rate approximately as $(\dot\gamma \tau)^{1/2}$ where the time scale $\tau$ grows on approach to $\phi_c$ as $|\phi-\phi_c|^{-4}$.  This represents the first experimental measurement of the full set of shear rheology exponents on both sides of the jamming transition.  The observed scaling of stress as power laws of the ``distances'' $|\phi-\phi_c|$ and $\dot\gamma$ to point-J in the jamming phase diagram supports the notion that jamming is similar to criticality at phase transitions but in a nonequilibrium system.

\begin{acknowledgments}
This work was supported by the National Science Foundation through grants MRSEC/DMR05-20020 and DMR-0704147. We thank B. Polak for preliminary experiments; A. Alsayed for helping synthesize the particles; and A. J. Liu, T. C. Lubensky, B. P. Tighe, S. Teitel, and W. van~Saarloos for helpful discussions.
\end{acknowledgments}

\bibliography{RheologyRefs}

\end{document}